\documentclass[prl,twocolumn,showpacs,preprintnumbers,amsmath,amssymb]{revtex4}
\usepackage{txfonts}
\usepackage{epsfig,amsmath,amssymb} 
\usepackage[all]{xy}
\usepackage{ifpdf} 
\usepackage{graphicx}   % for figures
\usepackage{amsfonts}%for tick and cross 
\usepackage{makeidx}  
\usepackage{epigraph}
\usepackage{etoolbox}
\makeatletter
\patchcmd{\epigraph}{\@epitext{#1}}{\itshape\@epitext{#1}}{}{}
\makeatother

\newcommand{\qed}{\hfill \mbox{\raggedright \rule{.07in}{.1in}}}

\newcommand{\ket}[1]{\left | #1 \right\rangle}
\newcommand{\bra}[1]{\left \langle #1 \right |}

\newcommand{\qq}[1]{``#1"}

\begin{document}

\title{Witnessing quantumness of a system by observing only its classical features} 

\author{C. Marletto$^{a}$ and V. Vedral $^{a,b}$
	\\ {\small $^{a}$ Physics Department, University of Oxford} 
	\\{\small $^{b}$Centre for Quantum Technologies, National University of Singapore}}

\date{April 2016}

\begin{abstract}
	
\noindent Witnessing non-classicality in the gravitational field has been claimed to be practically impossible. This constitutes a deep problem, which has even lead some researchers to question whether gravity should be quantised, due to the weakness of quantum effects. To counteract these claims, we propose a thought experiment that witnesses non-classicality of a physical system by probing it with a qubit. Remarkably, this experiment does not require any quantum control of the system, involving only measuring a single classical observable on that system. In addition, our scheme does not even assume any specific dynamics. 
That non-classicality of a system can be established indirectly, by coupling it to a qubit, opens up the possibility that quantum gravitational effects could in fact be witnessed in the lab. 
\bigskip

\end{abstract}

\maketitle

Direct evidence in favour of the quantisation of gravity is, at present, hard to obtain. Despite the recent success in detecting gravitational waves, the detection of gravitons -- quantum particles mediating the gravitational field -- has been argued to be practically impossible \cite{NOB}, \cite{NOB1}. These impossibility claims lead one to question whether gravity should be quantised in the first place. 

Evidence for quantisation can also be gathered indirectly, by coupling the gravitational field to a quantum system. For example, Feynman \cite{FEY} considered a thought experiment where a test mass in a superposition of two different locations interacts with gravity (see figure 1).

% %%%%%%%%%%%%%%%%%%%%%%%%%%%%%%%%%%%%%%%%%%%%%%%%%%%%%%
\begin{figure}[h]
	\centering
	\includegraphics[scale=0.07]{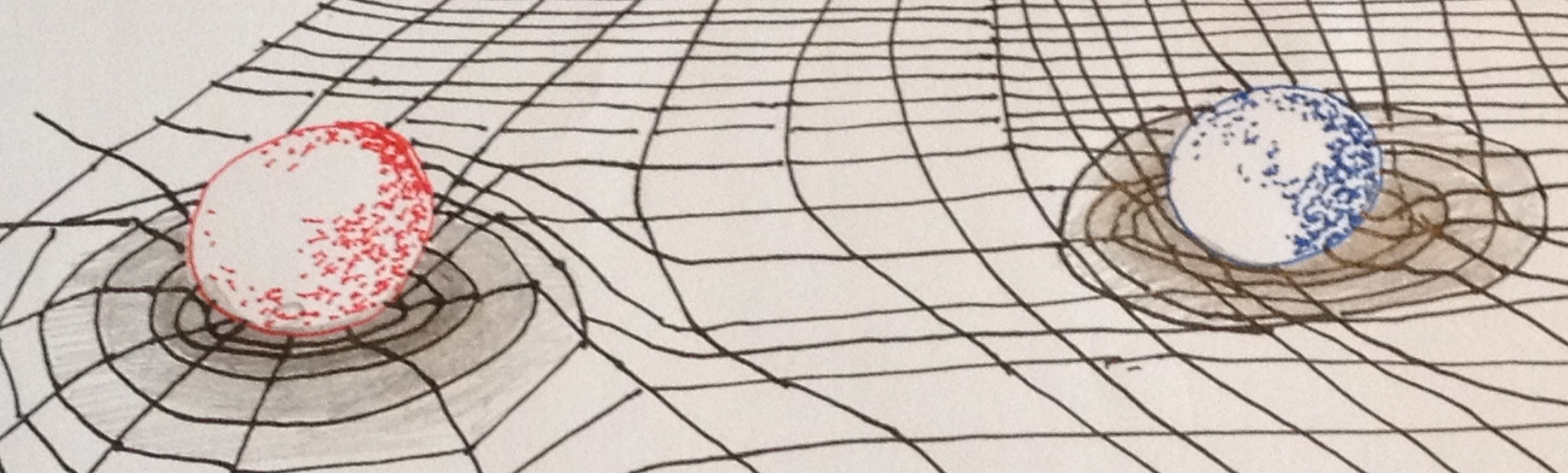} 
	\caption{An artist's impression of a mass in a superposition of two locations. In each of the two locations it interacts with gravity, thus creating correlations. Feynman's thought experiment explores the issue of  whether these correlations are classical or quantum. }
\end{figure}
%%%%%%%%%%%%%%%%%%%%%%%%%%%%%%%%%%%%%%%%%%%%%%%%%%%%

His point was that the physical state of the composite system would have different properties according to whether the mass is entangled with gravity, or just somehow classically correlated with it. These two situations could in principle be distinguished, but this requires to witness non-classicality in the gravitational field. Therefore, the thought experiment seems to conceal a circularity: witnessing non-classicality would seem either to require measuring two complementary observables on the field itself, or that the field undergoes some interference process (which itself frequently constitutes a measurement of two complementary observables). However, the possibility of performing either of those operations is precisely what the thought experiment is designed to assess. 

This is an instance of a more general problem, affecting the predictions of theoretical arguments in favour of quantisation. Such arguments  \cite{FEY, DeW, TER, HEI, BOHRING} claim that a subsystem of the universe (e.g. a field) interacting with a quantum system (which here means any physical system that can implement a qubit) must itself be quantised. The problem is, again, that testing these predictions might seem to require full quantum control of the system that is argued to be quantised; but the existence of such a control is what the test is supposed to probe.

In this paper we offer a solution to this problem. We propose a new thought-experiment to witness non-classicality of a system, by probing it via a qubit, without requiring any quantum control on the system. Specifically, the experiment is performed on the composite system of a qubit $S_Q$ and of a classical system $S_C$, which is assumed to have only a single observable $T$. By \qq{classical} we mean precisely a system that has only one single observable. Our proposed test for non-classicality only requires to measure correlations involving just the observable $T$ on $S_C$; and the system $S_C$ need not undergo any interference. Appropriate values of these correlations, as we shall show, imply that the classical system $S_C$ must have at least another observable that cannot be simultaneously sharp when the observable $T$ is. This is our indirect witness of non-classicality. We then conjecture the possibility of an indirect measurement of the complementary observable $S$, by coupling the classical system with a qubit, via a teleportation-type scheme. 

Remarkably, our thought-experiment is formulated in a general, information-theoretic framework -- which is independent of the details of the dynamics of the system $S_C$. Thus, our result is relevant within a wide range of different contexts, going well beyond quantum gravity: for example, it applies to testing non-classicality of macroscopic systems, be they biological systems \cite{DAV, DAVE} or computational devices \cite{TRO}. After all, in any experiment quantum control can only be assumed to exist on a limited number of degrees of freedom, while the rest could for all practical purposes be classical. 

\bigskip

Let $\hat q^{(1)}\doteq(\sigma_x\otimes I, \sigma_y\otimes I, \sigma_z\otimes I)$ denote the vector of generators $q_{\alpha}^{(1)}$ of the algebra of observables of the qubit $S_Q$, where $\sigma_{\alpha}, \alpha =x,y,z$, are the Pauli operators and $I$ is the single-qubit unit. Let $T$ be a binary observable on the classical system $S_C$ -- in other words, the classical system is supposed to be a single bit. Without loss of generality, we can represent it as an operator $q_z^{(2)}\doteq I\otimes\sigma_z$. For example, in the case of gravity, $T$ could be a discretised version of the position observable, representing two different locations of a mass which interacts with the quantum system through gravity. (If $S_C$ has a higher dimensionality, our result applies as long as one considers a quantum system $S_Q$ with the same dimensionality as $S_C$). 

Consider now an operation defined so that it performs a classical copy of the values held by $S_Q$ with $S_C$ as the target, in the basis defined by the observable $q_z^{(1)}q_z^{(2)}$. In other words, in that basis, it is required to perform the computation $\{00\rightarrow 00, 10\rightarrow 11\}$, where the first slot represents the value held by $S_Q$ and the second slot the value held by $S_C$. However, it is unknown what effect it has on other input states.  This is because our scheme is independent of the details of the dynamics. The thought experiment precisely investigates what happens when the input states are $\ket{\pm}\ket{0}$, where $q_1^{(z)}$ is not sharp. Those states therefore act as a probe. 

Note that the copy-operation is not assumed to be coherent, unitary or reversible in any sense. For example, it could be thought of as a classical controlled-NOT gate, where the NOT gate is realised by some classical evolution, such as $\ket{0}\bra{0}\rightarrow \cos^2( t)\ket{0}\bra{0}+\sin^2(t)\ket{1}\bra{1}$ for appropriate arguments $t$ (though the evolution need not be continuous).

 The thought experiment goes as follows. First, prepare the qubit $S_Q$ in the eigenstates $\ket{\pm}$ of $q_x^{(1)}$; and the classical system $S_C$ in some fixed state, which we denote as $\ket{0}$, representing a state where the observable $T$ is sharp with value $1$. Then, apply the copy-operation. Let us denote by $\rho_{\pm}$ the states of $S_Q\oplus S_C$ thus generated.  At this point, measure the averages $\langle q_{\alpha}^{(1)}q_z^{(2)}\rangle_{\pm}$, $\alpha=1,2,3$, and $\langle q_{z}^{(1)}\rangle_{\pm}$ on the states $\rho_{\pm}$. Note that the global states $\rho_{\pm}$ of the composite system cannot be argued to be entangled by construction. This is because $S_C$ need not obey quantum theory.

Now consider a different procedure, to prepare the states  $\tilde{\rho}_{\pm}$ by applying the same copy-operation, as above, on each of the states $\rho_{\pm}$. Then, measure $\langle A_i^{(1)}q_z^{(2)}\rangle_{\tilde \pm}$ for appropriate observables $A_i^{(1)}q_z^{(2)}$. See figure 2.

% %%%%%%%%%%%%%%%%%%%%%%%%%%%%%%%%%%%%%%%%%%%%%%%%%%%%%%
\begin{figure}[h]
	\centering
	\includegraphics[scale=0.4]{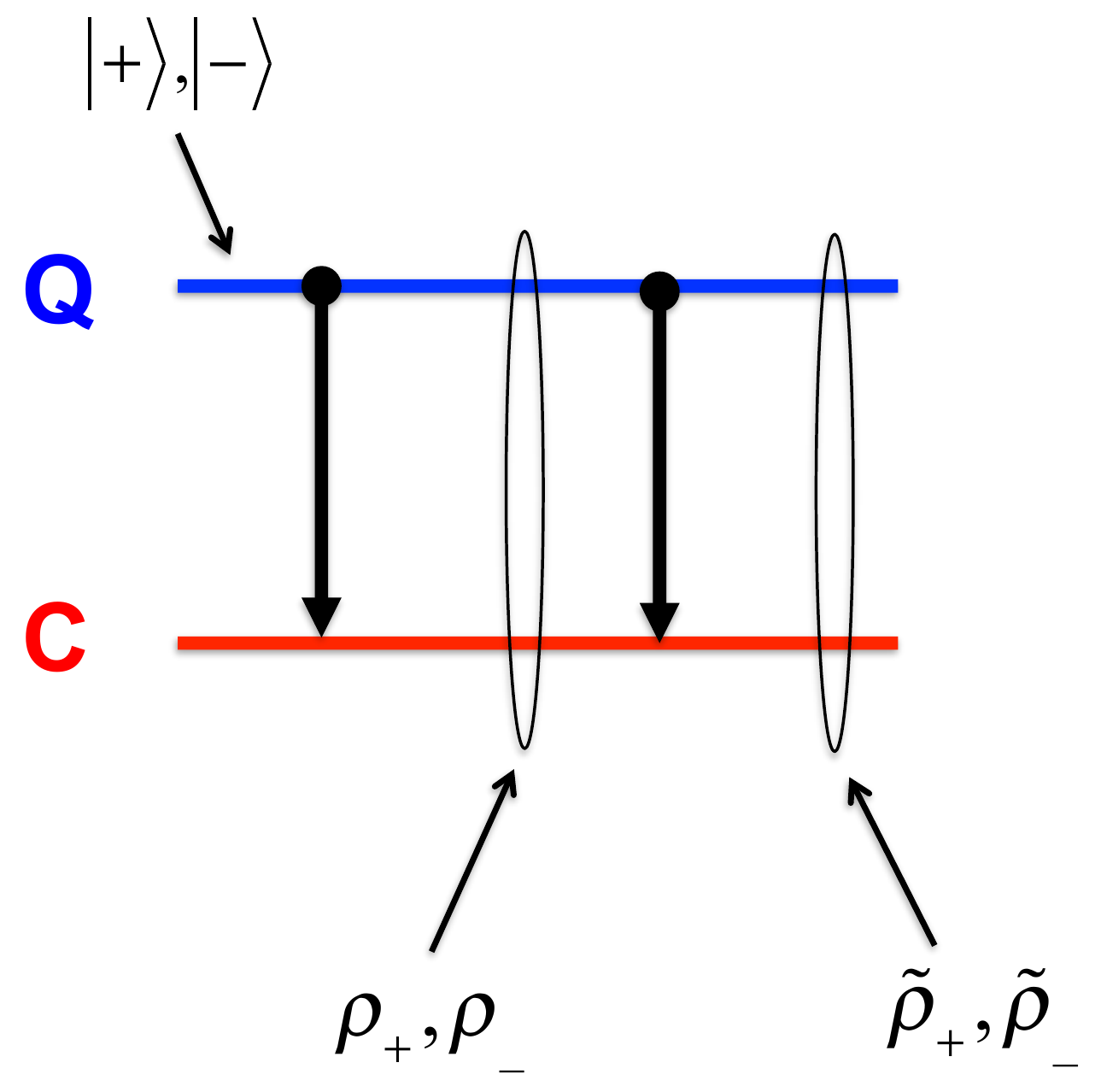} 
	\caption{Quantum network to prepare the states $\rho_{\pm}$ and ${\tilde\rho_{\pm}}$. $Q$ denotes the quantum system, $C$ the classical system. The black arrow represents a copier in the basis defined by  $q_z^{(1)}q_z^{(2)}$. No quantum control is required on $C$, as explained in the text.}
\end{figure}
%%%%%%%%%%%%%%%%%%%%%%%%%%%%%%%%%%%%%%%%%%%%%%%%%%%% 

We shall now argue that certain values of the correlation functions $\langle q_{\alpha}^{(1)}q_z^{(2)}\rangle_{\pm}$ and $\langle A_i^{(1)}q_z^{(2)}\rangle_{\tilde \pm}$ imply that the classical system must have at least another observable that is complementary to $T$. Crucially, both correlation functions only require the \qq{classical} observable $T$ to be measured on $S_C$.

Note first that, in our representation, the most general form of a state of $S_Q\oplus S_C$ is $$\rho = \frac{1}{4} \left ( I +\underline{r}.\hat q^{(1)}+s_z q_z^{(2)}+ \underline{t}.\hat q^{(1)} q_z^{(2)}\right)\;,$$
for some real-valued vectors $\underline r$, $\underline t$ and for some real coefficient $s_z$. This state, when interpreted as a two-qubit state, is separable and has no discord \cite{KAVAN}.

Now, suppose that $\langle q_z^{(1)}q_z^{(2)}\rangle_{\pm}=1$ and $\langle q_{\alpha}^{(1)}q_z^{(2)}\rangle_{\pm}=0$, $\forall \alpha\neq z$, are observed, for both $\rho_{\pm}$. This confirms that the quantum and the classical system have undergone some interaction, because that value differs from the same correlation function evaluated for the initial states $\ket{\pm}\ket{0}$. Suppose also that $\rho_{\pm}$ are ensemble-distinguishable from eigenstates of $q_z^{(1)}q_z^{(2)}$, by measuring $\langle q_{\alpha}^{(1)}\rangle_{\pm}$. This rules out the possibility that $\rho_{\pm}$ are themselves eigenstates in that basis. To satisfy these conditions, one must require $\underline{r}=\underline{0}$, $s_z=0$ and $\underline{t}=(0,0,1)$. Thus: $\rho_+=\rho_-=\frac{1}{4} \left ( I + q_z^{(1)} q_z^{(2)}\right)\;.$ 

Suppose further that it is possible to find observables $A_i$ of the qubit with the property that measuring $\langle A_i^{(1)}q_z^{(2)}\rangle_{\tilde\pm}$ can distinguish $\rho_{_{\tilde\pm}}$. This implies that $\rho_+\neq\rho_-$, which is a contradiction. 

Hence, we conclude that in order to reproduce the above correlation functions, the classical system must have an additional observable $T'$ that cannot be simultaneously sharp when $T$ is. In our representation,  that observable can be represented as an operator $q_x^{(2)}$ which does not commute with $q_z^{(2)}$. Our thought experiment therefore constitutes a witness of non-classicality on the system $S_C$, as promised. 

This non-classicality could be more general than the strictly quantum one. For example, $T$ and $T'$ might be two overlapping distributions in phase space, corresponding to uncertainty in preparation, as in Spekkens' toy model, \cite{SPEK}. This would effectively correspond to $S_C$ consisting of two classical bits whose values cannot be perfectly resolved. However, this model does not have a natural dynamics, so it is unclear what it would imply as to the physics of this thought experiment, or about the physical constitution of $S_C$. 

Crucially, our test only requires applying the copy-operation in the basis defined by $T$, in order to prepare the relevant states; and that those states can be discriminated by {ensemble measurements}, realised by measuring the local observables $A_i$ on the qubit and just $T$ on the $S_C$. These states could, in particular, be not perfectly distinguishable via single-shot measurements (thus the overall evolution might be non-unitary). Therefore our thought experiment does not presuppose the possibility of performing any interference on the classical system; nor the possibility of measuring other observables than $T$.

Whilst this test allows one to conclude that there must be an additional observable $T'$ which is necessary to describe the accessible states of $S_C$, that observable need not be measurable directly. It might be, for example, that there is some fundamental limitation to how well its eigenstates can be resolved from one another: this highlights an interesting distinction between an observable being directly measurable and its being necessary to describe the accessible states of a system. 

We now discuss how assuming the possibility of a coherent interaction between the classical system and some other qubit allows one to measure indirectly that observable. The scheme goes as follows. 

As before, prepare the states $\rho_{\pm}$ on $S_Q\oplus S_C$. At this point, apply the operation which in the basis defined by $q_z^{(1)}q_z^{(2)}$ realises the computation $\{00\rightarrow 00, 01\rightarrow 11\}$. This is a copy of the $q_z$ values held by $S_C$, this time with $S_Q$ as the target. As explained in \cite{MAVE}, this operation must generate, acting on each of the states $\rho_{\pm}$, two new (possibly mixed) states ${\alpha_+}$ and ${\alpha_-}$ on $S_C$.  As we said, these states need not be distinguishable from one another; moreover, since it is not possible to measure any other observable than $T$ on $S_C$, one cannot reconstruct them by applying a procedure such as state tomography directly to $S_C$. However, by bringing in another qubit $S_Q'$, it is possible to apply on $S_C\oplus S_Q'$ a sequence of three CNOT gates in the $q_z$ basis to perform a logical swap \cite{NIE}. This allows one to prepare the qubit $S_Q'$ in each of those two states. At this point, state tomography on the qubit allows one to distinguish the states asymptotically, thus showing indirectly that those states existed on $S_C$. See figure 3.

% %%%%%%%%%%%%%%%%%%%%%%%%%%%%%%%%%%%%%%%%%%%%%%%%%%%%%%
\begin{figure}[h]
	\centering
	\includegraphics[scale=0.4]{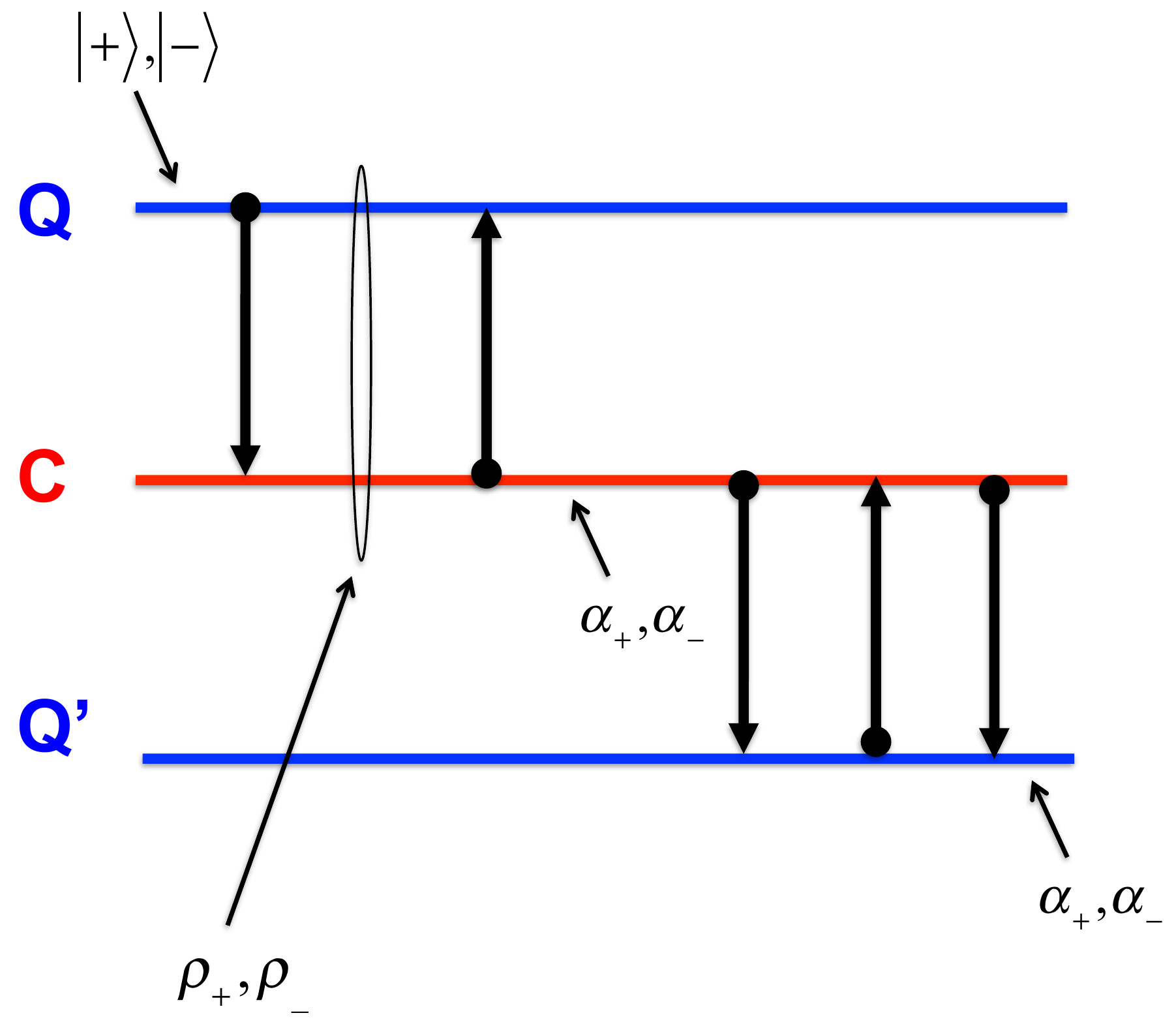} 
	\caption{Quantum network to prepare the states $\alpha_{\pm}$ on the classical system $C$. $Q'$ represents the second qubit where the states $\alpha_{\pm}$ are accessible. Again, no quantum control is required on the system $C$ (see explanation in the text).}
\end{figure}
%%%%%%%%%%%%%%%%%%%%%%%%%%%%%%%%%%%%%%%%%%%%%%%%%%%% 

Incidentally, if the states ${\alpha_+}$ and ${\alpha_-}$ are not orthogonal to one another, an orthogonalisation procedure would allow one to argue for the existence of {\sl two} other observables of $S_C$, in addition to $T$. Each one observable would be represented by an operator having, respectively, ${\alpha_+}$ and ${\alpha_-}$ as eigenstates. In the case of this test, too, the only observable measured on the classical system is $T$; however, overall coherence is required to realise the swap.

We emphasise that the central feature of our analysis is that it does not assume any particular dynamical model for the system whose non-classicality is to be witnessed, nor for the coupling between it and the quantum system.  This is in contrast to the recent argument in \cite{tom} where quantumness of a given system is confirmed indirectly by its ability to entangle two other bona fide quantum systems. In that sense the argument in \cite{tom} is an extension and elaboration of Feynman's argument \cite{FEY}. In the latter the witness of non-classicality would be the existence of bipartite entangled states generated by coherent evolution, while in \cite{tom} the witness is the existence of states with discord between the classical and the quantum systems, which forces the classical system to have a complementary observable. In both these arguments, however, the interactions are all assumed to be unitary and furthermore the system to be tested is described within the same formalism; whereas in our scheme, crucially, there are no such assumptions. 

Our thought experiment is also of practical importance as it illustrates how to manipulate quantumly a system on which there is no direct quantum control. Thus, our thought experiment is related to the experiments that prepare and subsequently witness  \qq{Schr\"odinger cat} states of light, \cite{HAR}. There, the system $S_C$ corresponds to the EM field, initially prepared in a coherent state; $S_Q$ is an atom, initially prepared in a superposition of its two energy levels. The two systems interact in such a way that they evolve into an entangled state, which corresponds to one of our $ \rho_{\pm}$. By measuring the atom in a complementary basis, the field is left in one of two orthogonal {Schr\"odinger cat states}. The atom (or an ensemble of them) is then also used as a probe, to witness the {cat states}. In this way the complementary observables of the EM field are never measured directly, just like in our thought experiment. 

Witnessing non-classicality of a physical system that need not obey quantum theory is a key task in contemporary physics.  It is crucial for testing predictions that gravity is quantised; but also to explore the quantum-to-classical boundary. For example, it is necessary to test predictions that macroscopic systems (e.g. a bacterium) coupled to a quantum system are, themselves, quantum. In all such cases, one cannot assume full quantum control on the physical system $S_C$ whose non-classicality is to be witnessed. For instance, tests of non-classicality designed for quantum systems, e.g. violation of Bell inequalities, are inadequate. Our thought experiment is a proposal for a new approach to performing that task.  Its strength is that, by using a quantum probe, it provides an indirect witness of non-classicality, which requires only measuring a single observable on the system $S_C$. In addition, it is remarkably general: it only relies on information-theoretic witnesses, without assuming any particular dynamics for the system $S_C$. Thus, our experiment is applicable to all the above open problems -- a task that we leave for future work.

\textit{Acknowledgments}: 
The authors thank Tomek Paterek and Tanjung Krisnanda for helpful comments. CM's research was supported by the Templeton World Charity Foundation. VV thanks the Oxford Martin School, the John Templeton Foundation, the EPSRC (UK) and the Ministry of Manpower (Singapore). This research is also supported by the National Research Foundation, Prime Minister’s Office, Singapore, under its Competitive Research Programme (CRP Award No. NRF- CRP14-2014-02) and administered by Centre for Quantum Technologies, National University of Singapore.

\end{document}